\documentclass[aps,preprint]{revtex4}
\usepackage{graphicx}
\usepackage{amssymb}

\begin{document}
\title{Evidence for Nodal Quasiparticles in the Non-magnetic Superconductor YNi$_{2}$B$_{2}$C via Field-angle-dependent Heat Capacity} 
\author{Tuson Park}
\author{M. B. Salamon}
\affiliation{Department of Physics and Material Research Laboratory, University of\\
Illinois at Urbana-Champaign, IL 61801, USA}
\author{Eun Mi Choi}
\author{Heon Jung Kim}
\author{Sung-Ik Lee}
\affiliation{National Creative Research Initiative Center for Superconductivity and
Department of Physics, Pohang University of Science and Technology, Pohang\\
790-784, Republic of Korea}
\date{Dec 24, 2002}

\begin{abstract}
Field-angle dependent heat capacity of the non-magnetic borocarbide superconductor YNi$_{2}$B$_{2}$C reveals a clear fourfold oscillation, the first observation of its kind. The observed angular variations were analyzed as a function of magnetic field angle, field intensity and temperature to provide its origin. The quantitative agreement between experiment and theory strongly suggests that we are directly observing nodal quasiparticles generated along $<100>$ by the Doppler effect. The results demonstrate that field-angle heat capacity can be a powerful tool in probing the momentum-space gap structure in unconventional superconductors such as high T$_{c}$ cuprates, heavy fermion superconductors, etc.
\end{abstract}

\maketitle

Most superconductors behave \textquotedblleft conventionally;" electronic
excitations are suppressed by the BCS energy gap causing the electronic heat
capacity, for example, to be exponentially small at temperatures well below
the superconducting transition. Recently, a number of superconductors,
cuprates and heavy-fermion metals among them, have been found to be
\textquotedblleft unconventional" in that they exhibit gap-zero (or nodal)
points or lines in momentum space. Electronic excitations--nodal
quasiparticles (nqp)--are then observed at low temperatures, giving rise to
power-law rather than exponential, behavior. Unlike gapless
superconductivity, which can occur in conventional superconductors, the
Fermi momenta of these quasiparticles are restricted to nodal regions of the
Fermi surface, giving a strong directional dependence to various physical
properties. While pioneering work by Salamon and coworkers \cite{salamon,yu}
and by subsequent experimenters \cite{aubin,izawa} demonstrated the
directionality of thermally excited nqp's via the thermal conductivity,
direct detection has proven elusive.

The rare earth borocarbide superconductors RNi$_{2}$B$_{2}$C (R=Y, Lu, Tm,
Er, Ho, and Dy) had been generally thought to exhibit isotropic s-wave
pairing via electron-phonon coupling \cite{carter,mattheis,Michor,gonnelli,manolo}, but there is growing evidence that the gap function is highly anisotropic or has a nodal structure on the momentum surface \cite{nohara,boaknin,yang,izawa2,izawa,maki}. In specific heat measurements, a power law behavior was observed in the temperature
dependence and the electronic coefficient $\gamma (H),$ to the extent it can
be extracted, follows a square-root field dependence \cite{nohara,volovik}.
Low temperature thermal conductivity measurements showed low energy
excitations that also indicate unconventional superconductivity \cite{boaknin}. Recently, Izawa and coworkers reported a field-directional-angle
dependence of the thermal conductivity in YNi$_{2}$B$_{2}$C that suggested
point nodes along $<100>$ directions \cite{izawa}. However, their
interpretation is not definitive because the highly anisotropic nature of
the Fermi surface and nonlocality could lead to anisotropic macroscopic
phenomena even within the ab-plane \cite{Metlushko}. Angular specific heat
measurements, however, are insensitive to the Fermi surface topology and
reflect the gap anisotropy through the Doppler energy shift \cite{vekhter}. We report here the first observation of Doppler-induced angular dependence of the
electronic heat capacity in the non-magnetic borocarbide superconductor
YNi$_{2}$B$_{2}$C.

In order to study the elusive angular variation in the low temperature
specific heat \cite{moler,wang}, we built a probe specifically designed to look at the DOS variation with field angle. The probe was designed to fit into a Quantum
Design Physical Properties Measurement System with a 7~T transverse magnet. The heat capacity of YNi$_{2}$B$_{2}$C was measured by ac calorimetry in which the sample was heated by an
electronically modulated laser pulse and the induced ac temperature was
measured by chromel-gold/iron (0.07\%) thermocouple. In order to account for
the magnetic field dependence of the thermocouple wire, the ac signal from a
non-magnetic dummy sample was measured vs magnetic field in the temperature
range of interest. As the heat capacity is field independent, any change
with field arises from the thermocouple wire and was used to correct the
data. The details of the ac technique can be found elsewhere \cite{tuson}.

The single crystal YNi$_{2}$B$_{2}$C was grown by the high temperature flux
method using Ni$_{2}$B as a solvent. The details are described elsewhere 
\cite{cho,mun}. The inset of the Fig.~1 shows the superconducting transition at 0~T. The sharp thermodynamic transition ($\Delta T/T_{c} \leq 0.02$) indicates that the sample is homogeneous.

The main graph of Fig.~1 shows the field dependence of the heat capacity at
2.5~K. The sample was zero-field cooled to 2.5~K and a magnetic field was
applied along the $(100)$ direction. The open circles describe the heat
capacity with increasing field and the crosses, with decreasing field. There
is a deviation between the two below 1~T, which indicates that flux pinning
is important at low magnetic fields. The system enters into the normal state
above 5.2~T along this direction and the lower critical field is less than
0.1~T. The heat capacity with increasing magnetic field was fitted by $%
C_{0}+b$ $(H-H_{0})^{1/2}$ ($\mathbf{dashed\ line}$). The excellent square-root fit
up to the upper critical field indicates that YNi$_{2}$B$_{2}$C is a superconductor with an unconventional pairing symmetry. In an intermediate field range, $H_{c1}\ll H\ll H_{c2}$, the square root field dependence comes mostly from nqp's induced by the Doppler effect; the core-state contribution
becomes important as the field approaches $H_{c2}$ \cite{volovik,machida}. 

The upper panel of Fig.~2 shows the field-directional angular dependence of
the total heat capacity at 2~K in 1~T. The field angle $\alpha$ was measured with respect to the $a$-axis. The transverse magnetic field was rotated within the basal plane of YNi$_{2}$B$_{2}$C by increments of $3^{\circ }$ by a computer controlled stepping motor. Repositioning the sample by $33^{\circ }$ relative to the thermocouples and 
apparatus shifts the 4-fold pattern, but not the 2-fold contribution, by 
that angle. The total heat capacity consists of constant, 2-fold, and 4-fold contributions: $C_{total}(\alpha)=C_{0}+C_{2}(\alpha)+C_{4}(\alpha)$. The
field-independent constant $C_{0}$ is due to nonmagnetic contributions such
as the lattice heat capacity and thermally excited nqp's, and is determined
experimentally in C vs H data. The 2-fold contribution $C_{2}(\alpha )$ comes
from our experimental setup and has a functional form of $c_{2}\cos 2\alpha $. The Au/Fe thermocouple wire is a major source of this contribution but
misalignment of the basal plane of the sample against the field direction
could also lead to the 2-fold component because of the anisotropy between
in-field ab-plane and c-axis heat capacities. The dashed line in the upper
panel of the Fig.~2 is $C_{2}(\alpha )$; the 2-fold signal is about 48~$\%$ of
the 4-fold component at 1~T and increases with magnetic field up to 55~$\%$ at
4~T. The circles in the lower panel of the Fig.~2 describe the 4-fold part $C_{4}(\alpha)$, which clearly reveals the angular variation.

In the mixed state of an unconventional superconductor, the supercurrent
flows around a vortex lead to a Doppler energy shift, $\delta \omega \sim 
\mathbf{v}_{s}\cdot \mathbf{k}_{F}$, where $\mathbf{v}_{s}$ is the velocity
of the superfluid and $\mathbf{k}_{F}$ is the Fermi momentum of nodal
quasiparticles. When the field direction is normal to the plane containing
nodes, the DOS is the average over the whole Fermi surface, leading to a
square-root field dependence \cite{volovik}. When the field is in the nodal
plane, however, the Doppler shift has a field-direction dependence as well, $%
\delta \omega \approx \frac{E_{h}}{\rho }\sin \beta \sin (\phi -\alpha )$ 
\cite{vekhter}. Here $\phi $ is an azimuthal angle of the gap node and $%
\beta $ is a vortex current winding angle. E$_{h}=\frac{a \hbar v_{F}}{2}\sqrt{\pi
H/\Phi _{0}}$ is the energy scale associated with the Doppler effect. The
geometrical constant $a$ is order of unity, $v_{F}$ is the Fermi velocity,
and $\Phi _{0}$ is the flux quantum. The variable $%
\rho =r/R$, where $r$ is the distance from the vortex core and $2R$ is the intervortex distance. Working in the 2D limit, Vekhter $%
et.al. $ calculated the DOS , $N\approx (N_{1}+N_{2})/2$, for an in-plane
magnetic field when $\omega ,E_{h}\ll \Delta _{0}$ and for four nodes at
angles $\alpha _{n}$ from orthogonal axes in a plane : 
\begin{equation}
\frac{N_{i}(\omega ,h,\alpha )}{N_{0}}=\left\{ 
\begin{array}{ll}
\frac{\omega }{\Delta _{0}}(1+\frac{1}{2x^{2}}) & (x=\omega /E_{i}\geq 1) \\ 
\frac{E_{i}}{\pi \Delta _{0}x}[(1+2x^{2})\arcsin x+3x\sqrt{1-x^{2}}] & (x\leq
1),
\end{array}
\right.
\end{equation}
where i~=~1,2, $E_{1}=E_{h}|\sin (\alpha _{n}-\alpha )|$ and $E_{2}=E_{h}|\cos
(\alpha _{n}-\alpha )|$. $N_{0}$ is the normal-state density of states. The directional dependence of the Doppler effect, therefore, leads to the oscillation of the DOS where it has minima along nodes and maxima along antinodes. At T~=~0~K, the
oscillation is cusped with a contrast of $1/\sqrt{2}$ between the minima and
the maxima and is independent of the magnetic field. At a finite
temperature, however, the sharp contrast is washed out and the DOS oscillation now depends on the magnetic field.

As seen in Fig.~2, our data are well described by the cusped function $|\sin
2\alpha |$ which will be used in the analysis of the 4-fold variation. To make a quantitative analysis, we fit our data to
$C_{4}(h,\alpha )=c_{4}(1+\Gamma |\sin 2\alpha |)$. The coefficient $c_{4}$ and the angular contrast $\Gamma $ were treated as field dependent fitting parameters. The solid
line in the lower panel of Fig. 2 shows the fit with $\Gamma =0.05$ in 1 T.
The sharp minima along $<100>$ indicate that there exist gap minima or nodal structures along those directions. That the nodal positions are at $<100>$ is consistent with the angular thermal conductivity measurement \cite{izawa}.

The contrast $\Gamma $ depends on the gap geometry and dimensionality of the
superconductor. The Vekhter et al. model above assumes the Fermi momentum
of nqp's to be restricted to the nodal plane. Recent band calculations show
that YNi$_{2}$B$_{2}$C has a 3D electronic structure in spite of its layered
structure \cite{mattheis}. A more realistic calculation, therefore, is to
allow nqp Fermi momentum out of the nodal planes, which decreases the
amplitude of the angular variation. Following Won et al. \cite{won}, we use a modulated cylindrical Fermi surface and account for the 3D effect
by replacing the Doppler related energies E$_{1}$ and E$_{2}$ by $%
E_{1}^{3D}=E_{h}[\sin ^{2}(\alpha _{n}-\alpha )+\cos ^{2}\theta ]^{1/2}$ and 
$E_{2}^{3D}=E_{h}[\cos ^{2}(\alpha _{n}-\alpha )+\cos ^{2}\theta ]^{1/2}$
respectively and integrate the DOS over polar angle $\theta $: 
\begin{equation}
N(w,h,\alpha )=\frac{1}{2\pi }\int_{0}^{2\pi }\frac{1}{2}[N_{1}(w,h,\alpha
,\theta )+N_{2}(w,h,\alpha ,\theta )]d\theta \text{.}
\end{equation}
The heat capacity ratio, $C(2K,h,\alpha )/C(2K,h,0)$, calculated from the
DOS, is displayed as a 3D surface vs magnetic field-angle $\alpha $~(x-axis)
and $H$~(y-axis) in Fig.~3(a). In Fig.~3(b), we show the experimentally
obtained data sets after subtraction of the background contributions. Since
the in-plane oscillation is due only to the Doppler effect, $%
C_{4}(2K,h,\alpha )/C_{4}(2K,h,0)$ corresponds to the theory. We used 0.6, 1,
1.5, and 2~T data sets to construct the contour both in numerical and
experimental plots. The Doppler energy scale $E_{h}$ was adjusted to get a best fit, giving the Fermi velocity of $v_{F} = \sqrt{v_{a} v_{c}} = 1.3 \times 10^{7}$ cm/s which is comparable to other reported values \cite{kogan}. We used the geometrical constant $a=1$ and $\Delta = 1.76 k_{B}T_{c}$. Note that the angle contrast between experiment and theory are in good agreement suggesting that we are directly observing Doppler-shift generated nqp's. The periodicity of the peaks is $\pi /2$ and the oscillations grow with increasing field because of the increased Doppler energy.

Fig.~4 compares the contrast between node and antinode, $\Gamma
(2K,h)=[C(2K,h,\pi /4)-C(2K,h,0)]/C(2K,h,0)$ from experiment ($\mathbf{open\
circles}$) with numerical values ($\mathbf{dashed\ line}$). Above $h=0.5$, the data deviates from the theoretical prediction. One source of deviation is the Zeeman energy splitting $\mu H$, where $\mu $ is Bohr magneton, discussed
recently by Whelan and Carbotte \cite{whelan}. At a critical field $H_{c}$, Zeeman and Doppler energy scales become comparable and the DOS variation
with field angle vanishes. It is unlikely, however, that the deviation comes from the effect because the critical field is $\sim 10^{2}$~T. A more likely
possibility is that the deviation is evidence that the
core-state contribution to the electronic specific heat becomes significant
in the high field regime. When we include the core states, the parameter $%
C_{4}$ consists of the core and the nodal contributions : $C_{4}(\alpha
)=C_{core}+C_{node}(\alpha )$. Since the oscillation anisotropy $\Gamma $
measures the amplitude of the 4-fold pattern against the nodal part only, we
need to subtract the core part from the $C_{4}$, resulting in an increased
anisotropy in our data. In order for a quantitative analysis, it is essential
to distinguish the contributions from the core states and the extended states.

The inset of the Fig.~4 shows the contrast $\Gamma $ vs temperature in 1 T. The basal plane of the sample was misaligned by $26^{\circ }$ from the field direction, which produced a larger 2-fold contribution as was expected due to the heat capacity anisotropy between ab-plane and c-axis. The temperature dependence of the anisotropy was compared with the 3D nqp model that was used in the above analysis. The dashed line is the estimated anisotropy with the same Doppler energy scale with $v _{F}= 1.3\times 10^{7}$ cm/sec. It explains the temperature dependence of the oscillation amplitude very well, corroborating that the nodal quasiparticles are Doppler shifted.

Before concluding this report, we want to discuss the possible pairing symmetries of YNi$_{2}$B$_{2}$C. Observation of the four-fold angular variation in our heat capacity measurement limits the candidates to those with gap minima or nodes with singlet pairing states such as anisotropic s-wave, d-wave or mixtures of two or more order parameters. Recently, Izawa et al. reported a clear fourfold pattern in the c-axis thermal conductivity at 0.27 K, which sets the gap-anisotropy ratio to be much higher than 100 \cite{izawa}. One can still argue for gap minima, but a more natural explanation will be the presence of nodes in momentum space. We note that the field-angle heat capacity will have 4-fold pattern for both d-wave and anisotropic s-wave gap with nodes. Considering the significance of the pairing symmetry to understand the superconductivity in YNi$_{2}$B$_{2}$C and further to understand the interplay of magnetism and superconductivity in RNi$_{2}$B$_{2}$C (R = Tm, Er, Ho, and Dy), other measurements such as phase sensitive $\pi$ junction SQUID or tunneling measurements are required to distinguish those anisotropic order parameters

In summary, we have reported the first direct evidence for a variation of the density of states of nodal quasiparticles due to a field-induced Doppler energy shift through a fourfold field-angle oscillation in the heat capacity. The angular variation, together with the $\sqrt{H}$ variation of the heat capacity at 2.5 K, strongly suggests that the gap function is extremely anisotropic or, more likely, that YNi$_{2}$B$_{2}$C is an unconventional superconductor. The magnetic field-angle, field-intensity and temperature dependence of the heat capacity are in quantitative agreement with a model in which the fourfold pattern arises from Doppler-enhanced, fully 3D nodal quasiparticles with momenta in $<100>$ directions.

This project was supported in part by NSF Grant No. DMR 99-72087 and at Pohang by the Ministry of Science and Technology of Korea through the Creative Research Initiative Program. X-ray measurements were carried out in the Center for Microanalysis of Materials, University of Illinois, which is partially supported by the U.S Department of Energy under grant DEFG02-91-ER45439. We thank H. Yanagihara and C. D. Benson for help in building the probe. T. Park thanks N. Goldenfeld for stimulating comments and discussions.

\bibliography{ynbc}

\clearpage

\begin{figure}[tbp]
\centering  \includegraphics[width=16cm,clip]{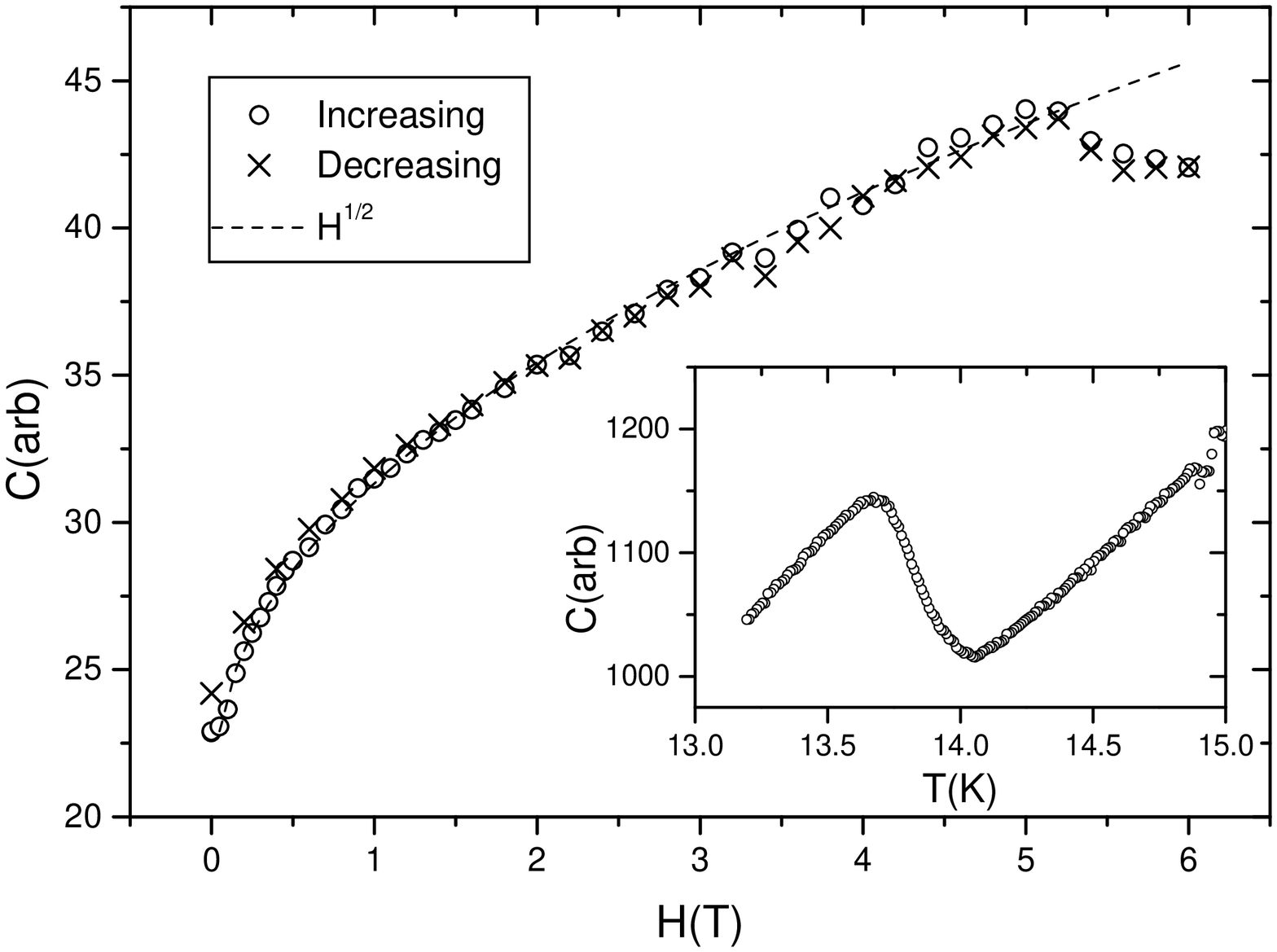}
\caption{Field dependence of the specific heat at 2.5 K along $(100)$ direction. The
open circles are with increasing magnetic field and the cross are with
decreasing field. The dashed line describes $C_{0}+b(H-H_{0})^{1/2}$ with $C_{0}=23$, $b=10$. $H_{0}$ is essentially due to Meissner effect and 0.09 T is used in the fit. The inset is the temperature dependence of the heat capacity
around superconducting temperature at zero field.}
\label{figure1}
\end{figure}

\begin{figure}[tbp]
\centering  \includegraphics[width=12cm,clip]{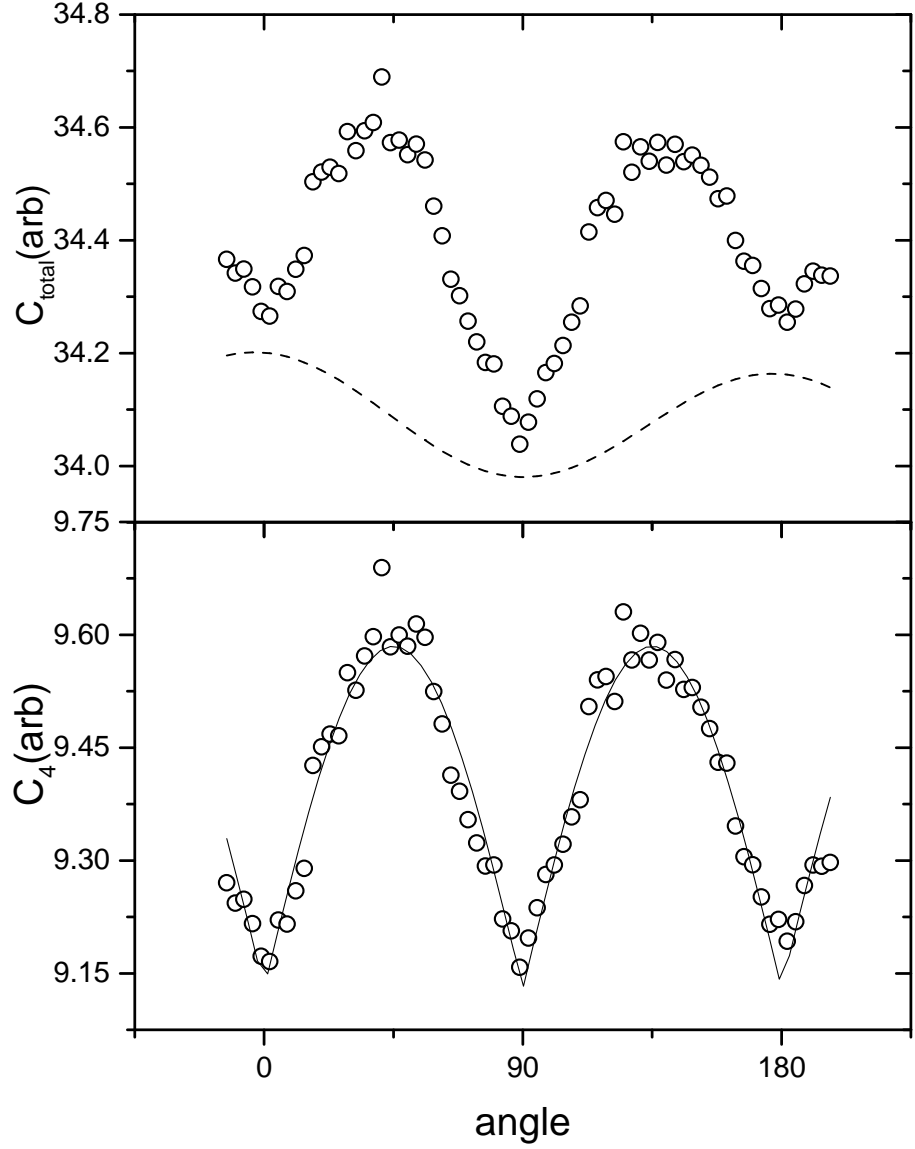}
\caption{Field directional dependence of the heat capacity at 2 K in 1 T.
The field angle $\protect\alpha$ is measured with respect to the a-axis. The
top panel shows the total heat capacity (open circles) and the 2-fold
component, $C_{2}(\protect\alpha)$ relative to baseline of 34.1 (dashed
line). The bottom panel shows the same data after subtracting the background, $C_{4}(\alpha)=C_{total}(\alpha)-C_{0}-C_{2}(\protect\alpha)$. The solid line
describes a fit with a cusped function, $C_{4}(\alpha)=c_{4}(1+\Gamma |\sin 2\protect\alpha|)$ with $\Gamma=0.05$.}
\label{figure2}
\end{figure}

\begin{figure}[tbp]
\centering  \includegraphics[width=10cm,clip]{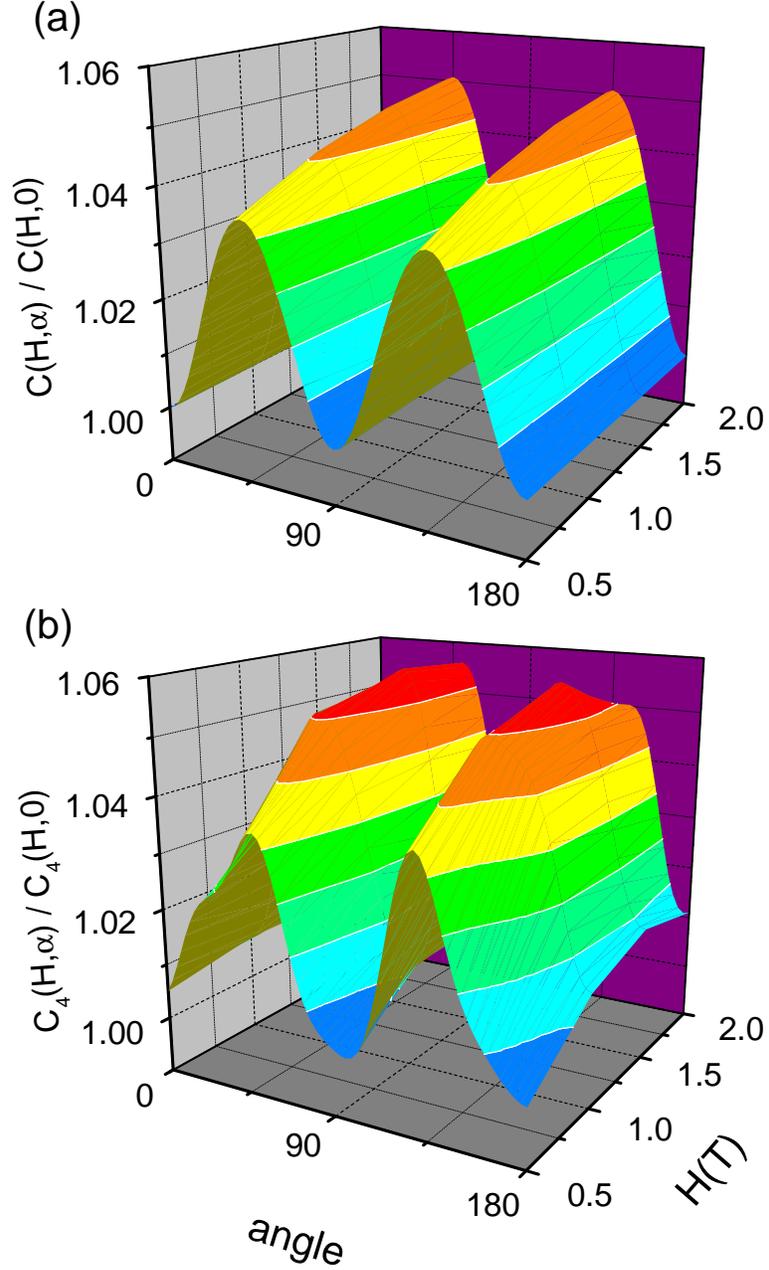}
\caption{The 3D surface plots describe magnetic field-angle (x-axis) and
field-intensity (y-axis) dependence of the in-plane specific heat ratio at 2~K: $C(h,\protect\alpha)/C(h,0)$. Fig.~(a) is a numerical calculation of the
specific heat due to nodal quasiparticles in 3D system and Fig.~(b) is
experimental data taken at 2 K in vortex phase. The experimental data were FFT low-pass filtered with a cut-off frequency of 0.03Hz with angles being considered as time. Both plots (a) and (b) were constructed with 0.6, 1, 1.5, 2 T data sets.}
\label{figure3}
\end{figure}

\begin{figure}[tbp]
\centering  \includegraphics[width=16cm,clip]{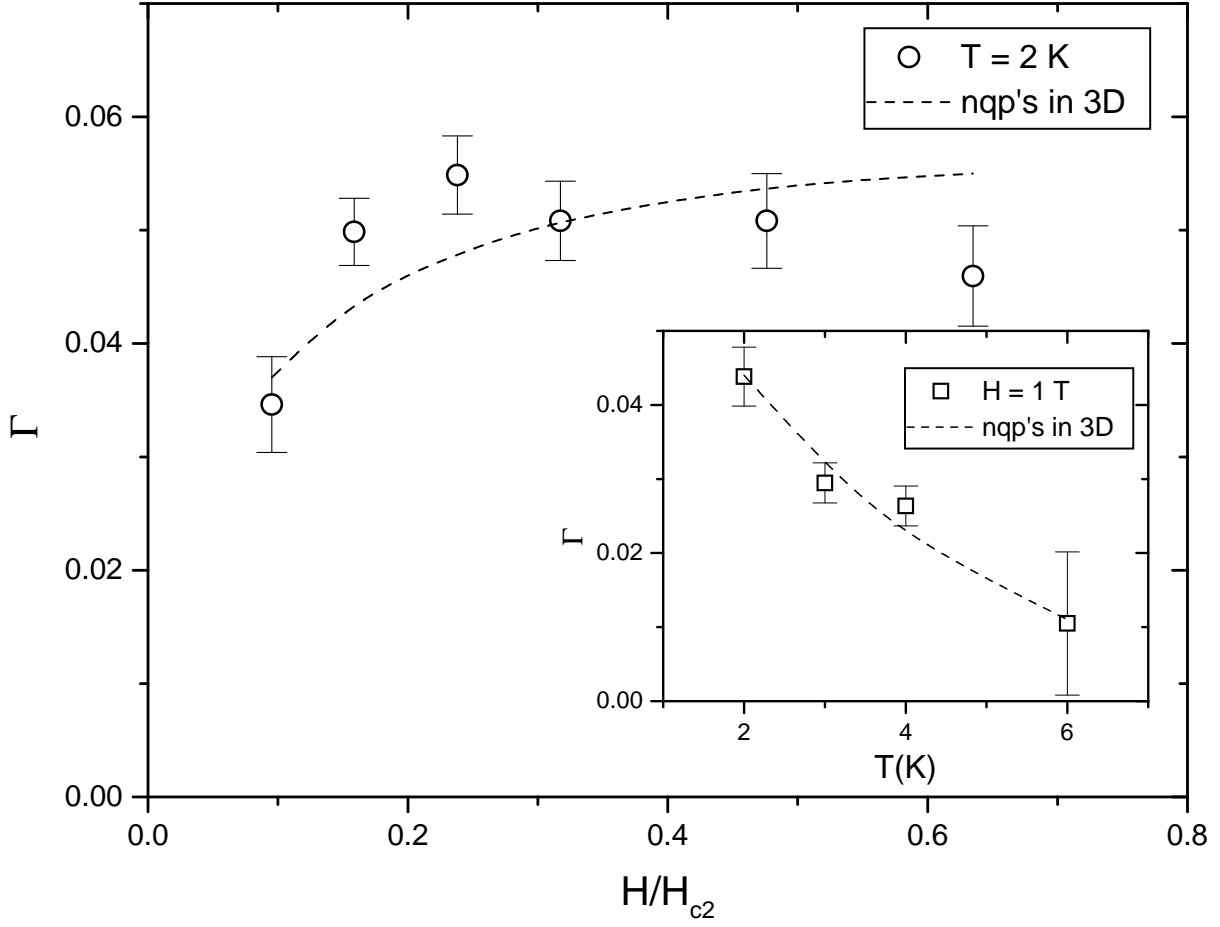}
\caption{The oscillation amplitude $\Gamma$ at $\protect\theta=90^{\circ}$
is shown against the reduced field, $H/H_{c2}$ at 2~K where $H_{c2}$
is 6.3~T. The dashed lines in the main graph and
in the inset are numerical estimations of the angular contrast by nqp's in
3D with the Fermi velocity $v_{F}=1.3\times 10^{7}$ cm/sec. Inset: The oscillation amplitude was shown against temperature in 1~T.}
\label{figure4}
\end{figure}

\end{document}